\documentclass[final,5p,times,twocolumn,authoryear]{elsarticle}

\makeatletter
\def\ps@pprintTitle{%
  \let\@oddhead\@empty
  \let\@evenhead\@empty
  \let\@oddfoot\@empty
  \let\@evenfoot\@oddfoot
}
\makeatother

\usepackage{amssymb}
\usepackage{lipsum}
\usepackage{color,soul}
\usepackage{flushend}

\begin{document}

\begin{frontmatter}

\title{Monte Carlo Experiments of Network Effects in Randomized Controlled Trials}

\author{Márton Trencséni (\texttt{mtrencseni@gmail.com})}

\begin{abstract}
This paper studies Monte Carlo simulations of content production in Randomized Controlled Trials on social networks modeled as random Watts-Strogatz graphs. Because treatment nodes are surrounded by control nodes, they do not receive the network boost they would receive if the entire network were treated, and part of their boost leaks into control. As a result, the treatment versus control lift underestimates the effect of a full graph-wide rollout. Clustering the treatment group reduces this underestimate, while its size does not depend on the size of the treatment group. These mechanisms appear as seven effects, each captured by a metric: the network effect, spillover effect, experiment dampening effect, intrinsic dampening effect, clustering effect, degree distribution effect and the experiment size effect.
 
\end{abstract}

\end{frontmatter}

\section{Introduction}

In academic as well as industry domains \textbf{Randomized Controlled Trials (RCTs)} are the gold standard tool for causal inference. Examples are clinical trials of new drugs and treatments, or improving online services and other online products (Kohavi, Tang and Xu, 2020). One of the core assumptions of traditional RCTs is the \textbf{Independence Assumption (IA)}, that the units of experimentation are independent of each other. In plain words, the assumption is that units don't communicate with each other, they do not affect other units' outcomes based on their own experiences --- whether they are in treatment or control. For many experiments ran on online services however --- social networks such as Facebook or Twitter/X --- the IA does not hold. The simplest toy example considered in this paper is a scenario of a treatment applied to a group of users that is intended to increase their content production. Increased content production of treatment users may boost content production of control users, since they will see their treatment friend's additional posts in their feed, and thus be motivated or inclined to themselves post more. This paper examines various, potentially unexpected effects that affect (contaminate) an experiment run on a network where the IA does not hold and defines metrics to measure the strength of these effects.

\section{Related work}

Experiments where one unit's treatment affects other units' outcomes are said to have \textit{interference}; standard references for causal inference under interference are Hudgens and Halloran (2008) and Aronow and Samii (2017). For experiments on social networks, Ugander et al. (2013) introduced graph cluster randomization, which assigns treatment to clusters of connected nodes, and Eckles, Karrer and Ugander (2017) showed that it reduces the bias of the difference-in-means estimator relative to the effect of treating the whole network. These correspond to the clustering and experiment dampening effects in this paper. The content production model used here is a linear peer effect model, whose steady state is the Katz-Bonacich centrality of the graph (Katz, 1953; Ballester, Calv\'o-Armengol and Zenou, 2006). Kohavi, Tang and Xu (2020) cover A/B testing in practice, and Albert and Barab\'asi (2002) review random graph models.

\section{Effects considered in this paper}

To discuss and understand these network effects, the paper uses Monte Carlo simulations of RCTs of content production over random Watts-Strogatz graphs. The effects considered and discussed in this paper are:

\begin{itemize}
\setlength\itemsep{0em}
    \item \textbf{Network effect}: each node's neighbours' content production boosts the node's own content production.
    \item \textbf{Degree distribution effect}: the network effect boost is a function of the graph's degree distribution.
    \item \textbf{Spillover effect}: in an experiment that boosts intrinsic content production in a subset of the nodes (the \textit{treatment group}), this boost will spill over to the rest of the nodes.
    \item \textbf{Intrinsic dampening effect}: the full lift of the treatment group is not realized in the network due to content production in the rest of the network being lower.
    \item \textbf{Experiment dampening effect}: we underestimate the lift in the treatment group as compared to control, due to the spillover effect and the intrinsic dampening effect.
    \item \textbf{Clustering effect}: a more tightly clustered treatment group leads to higher lift compared to control.
    \item \textbf{Experiment size effect}: as we add more nodes to the treatment group, absolute content production increases in the treatment group (and the entire network) due to the network effects, but not relative to the control group.
\end{itemize}

\begin{figure}[h]
	\centering 
	\includegraphics[width=0.34\textwidth]{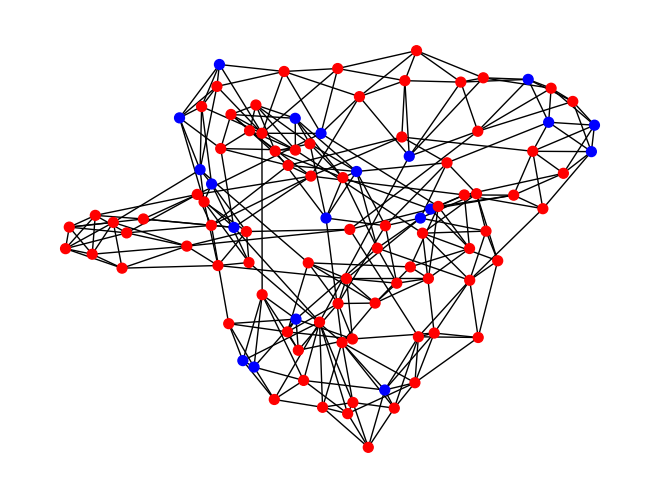}	
	\caption{Watts-Strogatz graph with $(n=100, k=6, p=0.1)$, with $N=20$ random treatment nodes colored blue.} 
\end{figure}

\section{Models of content production}

The network is modeled as graph $G = (V, E)$, where $V$ are nodes and $E$ are edges. For simplicity we assume edges are bi-directional, in other words the modeled relationship between nodes (such as \textit{friendship}) is symmetric. Each node $i$ has an associated random variable $R_i$, which describes the node's propensity for content production. Furthermore, we assume the content production random variables $R_i$ follow the same distribution $R$ and are parameterized by a variable $\lambda_i$, such that $R_i = R(\lambda_i)$. Without loss of generality, we will also assume that the random variables have an expectation value, and that $E[R_i] = E[R(\lambda_i)] = \mu_i$. Specifically, we will parameterize the distributions such that $E[R_i] = E[R(\lambda_i)] = \lambda_i$. Note that in some of the Monte Carlo simulations, in cases when we are investigating the mean behaviour, \textit{we will replace the random variable with its mean} --- after verifying this is justified.

We construct a dynamic model of $T$ discrete time steps: let $c_i^t$ be the amount of content produced by node $i$ at time step $t$. At $t=0$ all nodes have 0 content: $c_i^{t=0}=0$. In subsequent time steps, we use the random variables $R(\lambda_i)$ for each node $i$ to model content production, in other words $c_i^t \leftarrow R(\lambda_i^t)$, where $\leftarrow$ denotes drawing a value from the random variable. Note that we will omit indexes and superscripts at times to aid readability. We model the $\lambda_i$ as having two components $\lambda_i = \lambda_{int} + \nu_i$. The first parameter $\lambda_{int}$ models the node's \textit{intrinsic} propensity to produce content, independent of the rest of the graph, and is the same for all nodes on a network. When we model experiments, treatment nodes receive an additional intrinsic boost $\Delta \lambda$, like $\lambda_i = \lambda_{int} + \Delta \lambda + \nu_i$. The parameter $\nu_i$ models the propensity of the node to produce more content if its neighbours produce more content. We model $\nu_i$ as some function $f$ of the node's neighbours content production in the previous time step: $\nu_i^t = f(c_j^{t-1})$, where $j$ runs for all of $i$'s neighbours. Specifically, we will use $\nu_i^t = \nu_{damp} \sum_{j \in V_i}{c_j^{t-1}}$, where $V_i$ is the set of nodes that are $i$'s neighbours. $\nu_{damp}$ is an important global dampening parameter, whose presence is important to make the overall model stable in the steady-state (see next section).

The above model can be summed up as: at each time step $t$, for each node, we add up content produced by its neighbours in the previous time step, multiply by $\nu_{damp} $, add $\lambda_{int}$ (and $\Delta \lambda$ for treatment nodes), and draw with this parameter from the random variable $R$ to get content production at $t$.

Throughout this paper we will use connected Watts–Strogatz random graphs (Watts and Strogatz, 1998). A Watts–Strogatz graph is described by 3 parameters $(n, k, p)$. First, a regular ring lattice is constructed, a graph with $n$ nodes each connected to $k < n$ neighbors. Then, "shortcuts" are created by replacing some edges as follows: for each original edge $(u, v)$ in the graph, with probability $p$ replace it with a new edge $(u, w)$, where $w$ is randomly chosen. If the process yields a graph that is not connected, then try again until we get a connected graph. Note that the final randomization process does not change the mean degree in the graph, which remains $k$ at every step. Figure 1 shows a small example, with $N=20$ random treatment nodes.

\section{Mean behaviour, stability and the network effect}

In the model constructed, there are two sources of randomness: (i) randomness introduced when the graph is constructed, and (ii) the random variables used in content production (if one is used). As we will see, to understand the network effects discussed in this paper (ii) is not important, these random variables may be replaced by their means.

First, to examine the mean behaviour of the model we run simulations on a Watts-Strogatz graph with $(n=10\,000, k=50, p=0.1)$. We look at three models of content production: (i) a uniform distribution $R(\lambda)=U(0, 2 \lambda)$, (ii) a Poisson distribution $R(\lambda)=Pois(\lambda)$, and (iii) the mean constant case $R(\lambda)=\lambda$. Note that in all three cases, the mean is $E[R(\lambda)]=\lambda$.

Figure 2 shows that content production stabilizes after about $T=10$ steps, and that in all three cases mean content production across the network is the same, apart from random fluctuations coming from the random variables. Note that in case (i) and (iii) content production is modeled to be continuous, whereas in case of (ii) it's discrete, since the Poisson is a discrete random variable. Figure 3 shows the histogram of content production across nodes in the last time step, for all three models considered.

\begin{figure}[h]
	\centering 
	\includegraphics[width=0.34\textwidth]{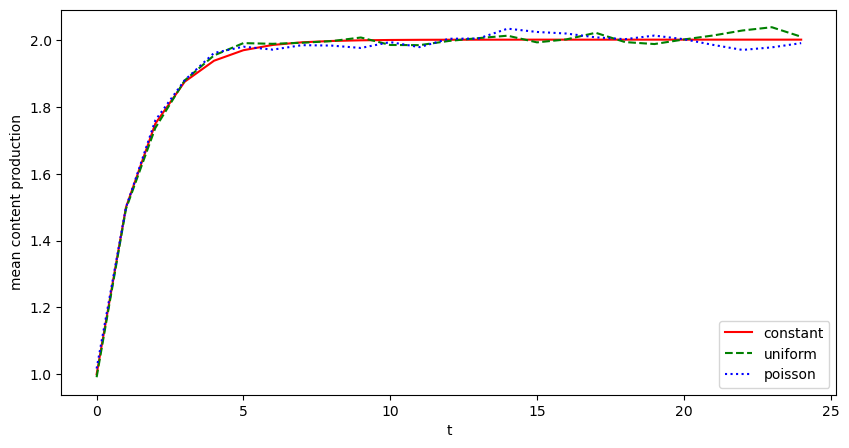}	
	\caption{Content production on Watts-Strogatz graph with $(n=10\,000, k=50, p=0.1)$, $\lambda_{int} = 1$ and $\nu_{damp} = 0.01$. Mean content production converges.} 
\end{figure}

\begin{figure}[h]
	\centering 
	\includegraphics[width=0.34\textwidth]{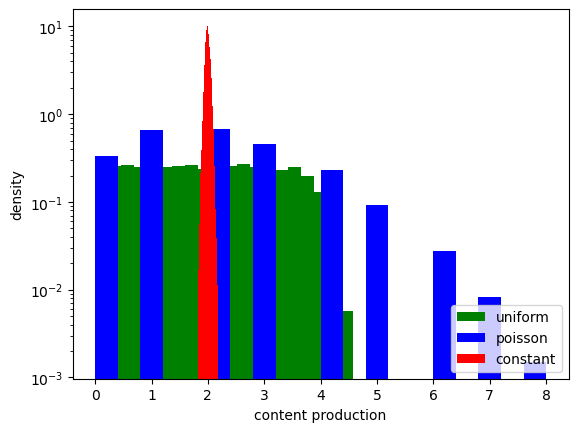}	
	\caption{Density histogram of content production during the last step of Monte Carlo simulation for the previous models. The y-axis is logarithmic to ease readability.} 
\end{figure}

\begin{figure}[h]
	\centering 
	\includegraphics[width=0.34\textwidth]{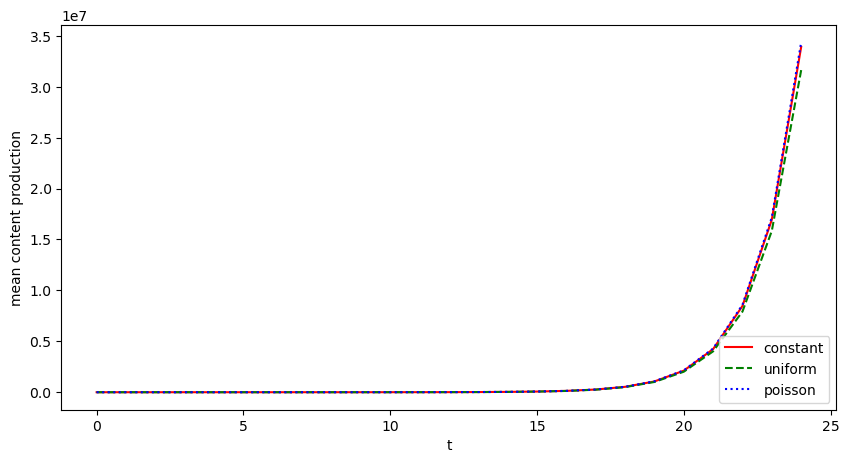}	
	\caption{Content production on Watts-Strogatz graph with $(n=10\,000, k=200, p=0.1)$, $\lambda_{int} = 1$ and $\nu_{damp} = 0.01$. Mean content production diverges.} 
\end{figure}

Next, let's re-run the same simulations, but on a Watts-Strogatz graph with $k=200$. The results are shown on Figure 4. With these parameters content production diverges, it does not reach a steady-state as before. Consider the mean constant case  to see why: assume that the model reaches a steady-state mean content production $c_{base}$. Then, from the way the model is constructed it follows that (using $E[R(\lambda)]=\lambda$):

\begin{equation}
    c_{base} = \lambda_{int} + c_{base} \cdot \bar{V_i} \cdot \nu_{damp}
\end{equation}

where $\bar{V_i}$ is the average number of neighbours per node, which for the Watts-Strogatz graph is $\bar{V_i} = k$. This equation is exact when every node has exactly $k$ neighbours ($p=0$); the next section discusses $p>0$. Solving for $c_{base}$ we get:

\begin{equation}
    c_{base} = \frac{\lambda_{int}}{ 1 - k \cdot \nu_{damp} }
\end{equation}

For the first model with $\lambda_{int} = 1, k = 50, \nu_{damp} = 0.01$ this yields $c_{base}=2$, which matches the numeric results. Clearly, this equation yields the stability condition:

\begin{equation}
    k \cdot \nu_{damp} < 1
\end{equation}

This is exact for $p=0$. For a general graph with adjacency matrix $A$, the steady state is $c = (I - \nu_{damp} A)^{-1} \lambda_{int}$, and the stability condition is $\nu_{damp} \rho(A) < 1$, where $\rho(A)$ is the largest eigenvalue of $A$ (Ballester, Calv\'o-Armengol and Zenou, 2006). Since $\rho(A) \geq k$, with equality only for regular graphs, $k \cdot \nu_{damp} < 1$ is necessary but not sufficient. For the Watts-Strogatz graphs in this paper, $\rho(A)$ is within $0.6$ of $k$, and within $0.2$ of $k$ for $p=0.1$.

For the divergent case, the above inequality is not satisfied. It is interesting to see the case where the denominator in Equation (2) is exactly 0, which we can achieve by setting $k=100$. In this case, content production does not explode, but increases approximately linearly over the simulated time window. Since this graph is not regular, $\nu_{damp} \rho(A)$ is slightly above 1, and growth eventually becomes exponential.

\begin{figure}[h]
	\centering 
	\includegraphics[width=0.34\textwidth]{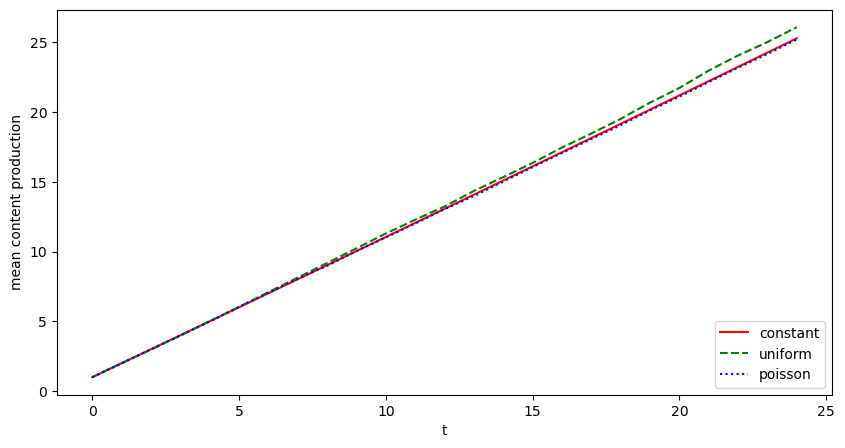}	
	\caption{Same as before, with $k=100$.} 
\end{figure}

This simple experiment of Figure 2 shows the \textbf{network effect}: we parameterized intrinsic content production with $\lambda_{int}=1$, which itself, without the boost from neighbours, would have yielded mean content production $c_{base} = \lambda_{int} = 1 $ in this simple model. However, due to the network effect, which here is a function of $k$ and $\nu_{damp}$, the actual mean content production $c_{base}$ is significantly higher, by a factor of $ \frac{ 1 }{ 1 - k \cdot \nu_{damp} } $.

\section{A note on regularizing models}

Real networks can experience exponential growth for a while, but not indefinitely --- some physical resource, such as people, attention or time runs out eventually. In this sense, the models above have unrealistic or un-physical domains because they can diverge. In principle, we can regularize the model in two places:

\begin{itemize}
\setlength\itemsep{0em}
    \item regularize the network effect: assume that each node can only consume a finite amount $\nu_{max}$ of its neighbours' content in a time step
    \item regularize content production: assume that each node can only produce a finite amount $c_{max}$ content
\end{itemize}

In both cases, the assumption is physical in a real network, because only finite time is available per time step to read or write content.

The simplest way to regularize is to cut off with the $min\{\}$ function, so that $\nu_i^t = min\{ \nu_{max}, \nu_{damp} \sum_{j \in V_i}{c_j^{t-1}} \}$ and $c_i^t \leftarrow min\{ c_{max}, R(\lambda_i^t) \}$. A smooth approach is to use a sigmoid function $\sigma(\cdot)$ to achieve smoother cut-off: $\nu_i^t = \nu_{max} \, \sigma \bigl( \nu_{damp} \sum_{j \in V_i}{c_j^{t-1}} \bigr) $ and $c_i^t \leftarrow c_{max} \, \sigma \bigl( R(\lambda_i^t) \bigr) $.

\begin{figure}[h]
	\centering 
	\includegraphics[width=0.34\textwidth]{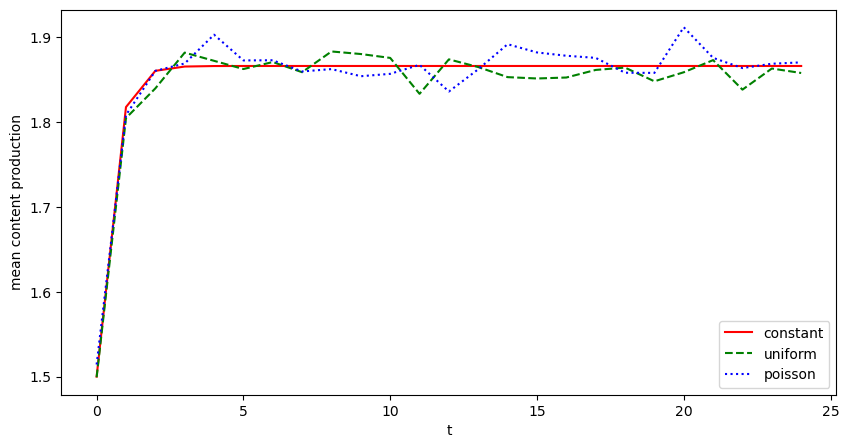}	
	\caption{Same as Figure 5, but with regularized network effect using the sigmoid $\sigma(\cdot)$ and $\nu_{max}=1$. Content production is no longer divergent.} 
\end{figure}

Figure 6 shows the same model as Figure 5, but with (just) the network effect regularized. In this case, in the mean-field approximation (exact for a regular graph with the mean constant model), solving for $c_{base}$ we get:

\begin{equation}
    c_{base} = \lambda_{int} + \nu_{max} \sigma( c_{base} \cdot \bar{V_i} \cdot \nu_{damp} )
\end{equation}

Due to the properties of $\sigma(\cdot)$, there is no closed form solution for $c_{base}$ in this case. In the rest of this paper, we will \textit{not} use regularization, as it makes it harder to reason about the behaviour of the network --- we will work with stable "raw" models.

\section{Degree distribution effect}

In this section we will discuss a small, nuanced adjustment to the above formula for mean content production on a random graph \textit{that is due to the degree distribution of nodes}. In a Watts-Strogatz graph with $p=0$, all nodes have exactly $k$ neighbours; however, at $p>0$, due to the random adjustment of edges, some nodes will end up with lower, and some with higher degree count than $k$. Figure 7 shows the degree distribution of a Watts-Strogatz graph with $k=50$ at different $p$ values. It shows that with increasing $p$, the distribution spreads out around the mean of $k$.

\begin{figure}[h]
	\centering 
	\includegraphics[width=0.34\textwidth]{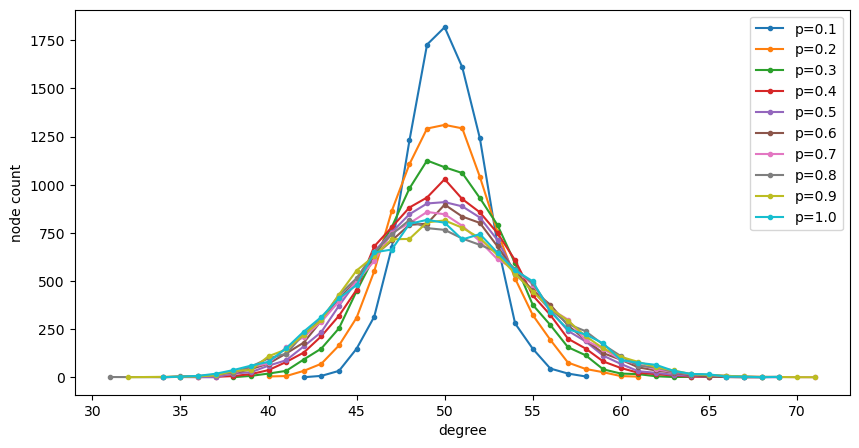}	
	\caption{Degree distribution of a Watts-Strogatz with $n=10\,000, k=50$ at different $p$ values.} 
\end{figure}

When examining content production on these random graphs at different $p$ values, an interesting effect shows itself: mean content production increases in a concave way over the base value, which is only realized exactly in the $p=0$ case. Defining

\begin{equation}
    e_{degree\,distribution} = \bar{c} / c_{base} - 1 
\end{equation}

where $\bar{c}$ is the actual content production measured in the network in steady-state and $c_{base}$ is the theoretically calculated value from Equation (2). Figure 8 shows the effect as a function of $p$.

\begin{figure}[h]
	\centering 
	\includegraphics[width=0.34\textwidth]{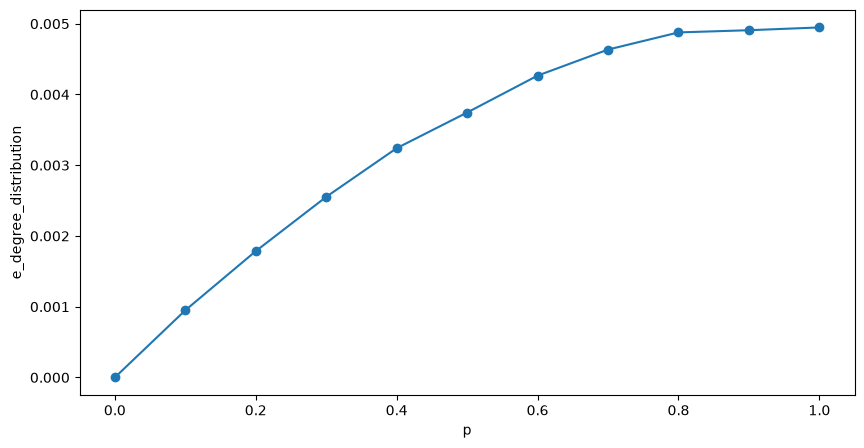}	
	\caption{Increased content production due to more spread out degree distribution of a Watts-Strogatz with $n=10\,000, k=50$ at different $p$ values.} 
\end{figure}

The explanation is as follows: the mean degree of nodes remains exactly $k$ in a Watts-Strogatz graph, despite (and during) the final randomization of edges, since the total number of nodes and edges always remains the same. However, as the histogram in Figure 7 shows, some nodes will end up with higher, and some with lower degree. Nodes with higher degree get a higher network boost and produce more content, and since they are also the neighbours of more nodes, their higher production is counted more often in their neighbours' network boost. The lower production of low-degree nodes does not cancel this out: averaging the steady-state equation over nodes gives $(1 - k \nu_{damp}) \bar{c} = \lambda_{int} + \nu_{damp} \mathrm{Cov}(d_i, c_i)$, where $d_i$ is the degree of node $i$, so mean content production increases with the covariance of degree and content production, which grows with the spread of the degree distribution. This is illustrated in Figure 9, which shows the mean content production for the same graph as above, but as a function of node degree.

\begin{figure}[h]
	\centering 
	\includegraphics[width=0.34\textwidth]{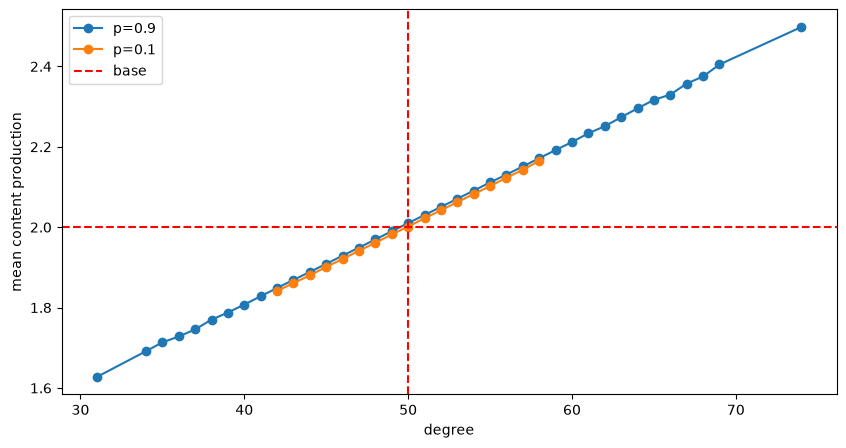}	
	\caption{Content production in the previous network, as a function of node degree. The horizontal red dashed line shows the baseline content production we would have without a spread of degree distribution, the vertical red dashed line is the mean degree. Clearly content production is a function of degree, and the spread is wider in the positive direction.} 
\end{figure}

In the rest of the paper we will use the primed notation $c_{base}\prime$ to refer to the "bare" content production on the network, taking into account the above effect (so slightly higher than $c_{base}$). Bare here means no treatment is applied, all nodes are identical. The steady state is $c = (I - \nu_{damp} A)^{-1} \lambda_{int} \mathbf{1}$, so $c_{base}\prime$ is $\lambda_{int}$ times the mean Katz-Bonacich centrality of the graph (Katz, 1953; Ballester, Calv\'o-Armengol and Zenou, 2006). In the rest of the paper it is calculated from a bare Monte Carlo simulation of a network with $n=10\,000$ nodes and the same $k$ and $p$.

\section{Spillover effect}

In this section we will start simulating \textbf{Randomized Controlled Trials (RCTs)} on networks. RCTs are traditionally run with the \textbf{Independence Assumption (IA)}, meaning the units of experimentation are independent of each other. In the network models considered here, the IA does not hold --- the resulting behaviour is the topic and \textit{raison d'etre} of this paper itself.

Consider a model on a Watts-Strogatz graph with $(n=500\,000, k=50, p=0.1)$, with identical parameters as before, without regularization. Per Equation (2), such a model has an expected steady-state content production of $c_{base}\prime \approxeq 2$. This base model is modified so that a random 2\% treatment subset of nodes receives a $\Delta \lambda = 0.05$ boost to their intrinsic content production. Figure 10 shows content production separately for \textbf{treatment}, \textbf{neighbours} of treatment and the \textbf{rest} of nodes (these 3 are distinct sets covering the whole graph).

\begin{figure}[h]
	\centering 
	\includegraphics[width=0.34\textwidth]{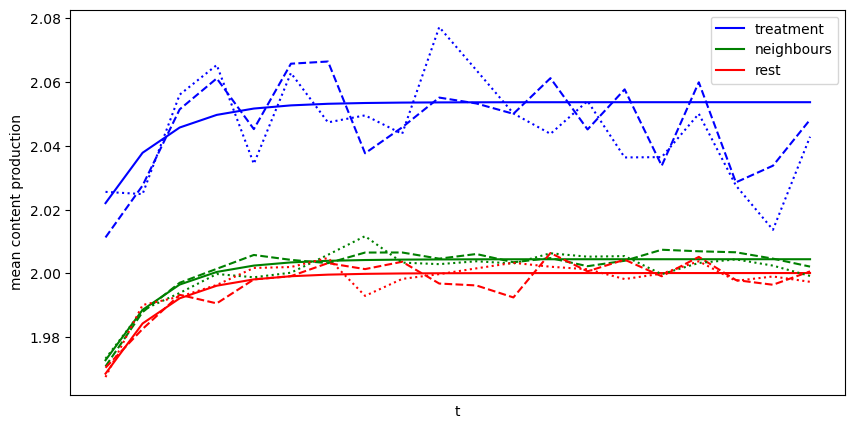}	
	\caption{Spillover effect in a Randomized Controlled Trial on a Watts-Strogatz graph with $(n=500\,000, k=50, p=0.1)$ with $N=10\,000$ treatment nodes receiving a $\Delta \lambda = 0.05$ boost to their intrinsic content production, for all three (mean constant, uniform, Poisson) content production models. The line types match the legend on previous figures. The first few time steps where steady-state is approached is not shown.} 
\end{figure}

The following table shows the split of nodes and the average content production $c$ level of the various subsets in the network (\textbf{control} is the union of \textbf{neighbours} and \textbf{rest}):

\begin{center}
\begin{tabular}{ |c|c|c| } 
 \hline
 \textbf{group} & \textbf{cardinality} & $c / c_{base}\prime$ \\ 
 \hline
 treatment   &   $10\,000$   &  1.0258 \\ 
 control     &  $490\,000$   &  1.0005 \\ 
 neighbours  &  $313\,507$   &  1.0012 \\ 
 rest        &  $176\,493$   &  0.9991 \\ 
 \hline
\end{tabular}
\end{center}

What this shows is that the treatment effect \textbf{spills over} into the control group. The spillover is stronger into direct neighbours of treatment, weaker for the rest of nodes. It's also worth noting that treatment, control and neighbours subsets have higher content production than the simulated $c_{base}\prime=2.0019$ base value, which shows that even a small treatment group can "contaminate" the entire network. However, rest (nodes that are not treatment and also not neighbours of treatment) has slightly lower ($c_{rest}/c_{base}\prime < 1$) content production --- how is this possible? The explanation is statistical: the nodes in rest are biased to have fewer neighbours, since a node with more neighbours is more likely to be adjacent to a treatment node, and this results in this subset's content production being slightly lower due to lower network effects. However, the overall conclusions are unaffected by this observation.

We can measure the strength of the spillover effect by defining:

\begin{equation}
e_{spillover} = c_{control} / c_{base}\prime  - 1
\end{equation}

In other words, $e_{spillover}$ measures the degree of contamination (due to the treatment effect applied to the treatment group). In this scenario we measure $e_{spillover} = 0.0005$.

Clearly, the spillover effect must be a function of $N/n$ and $k$:

\begin{itemize}
\setlength\itemsep{0em}
    \item as $N/n \rightarrow 0, e_{spillover} \rightarrow 0$, because the treatment group becomes insignificant in the overall network
    \item as $k \rightarrow 0, e_{spillover} \rightarrow 0$, because nodes have fewer neighbours and the model approaches a conventional RCT where the Independence Assumption holds
\end{itemize}

To illustrate the dependence on $N/n$ and $k$, let's look at the same experiment as above, but with $k=10$ and a 1\% treatment group of $N=5\,000$ nodes (Figure 11). In this scenario, where $c_{base}\prime=1.1112$, the mean steady-state of the network is:

\begin{center}
\begin{tabular}{ |c|c|c| } 
 \hline
 \textbf{group} & \textbf{cardinality} & $c / c_{base}\prime$ \\ 
 \hline
 treatment   &  $5\,000$    &  1.0450 \\ 
 control     &  $495\,000$  &  1.0000 \\ 
 neighbours  &  $47\,275$   &  1.0014 \\ 
 rest        &  $447\,725$  &  0.9999 \\ 
 \hline
\end{tabular}
\end{center}

The spillover effect works out to be $e_{spillover}=0.00005$, about 10x weaker than in the previous example.

\begin{figure}[h]
	\centering 
	\includegraphics[width=0.34\textwidth]{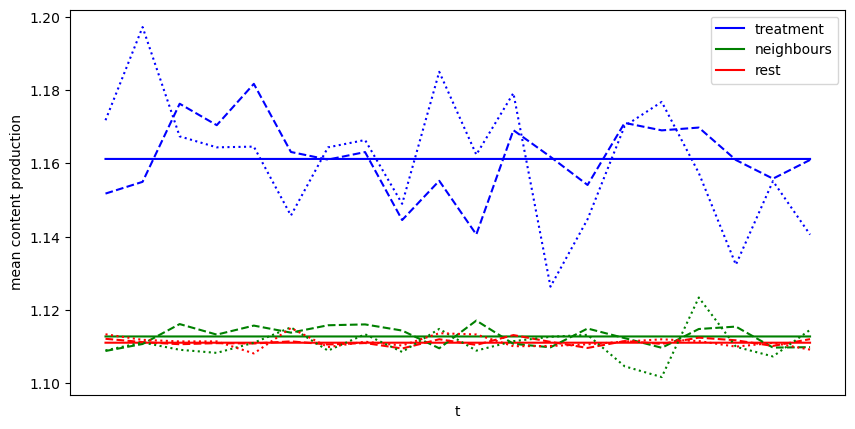}	
	\caption{Same as Figure 10, but with $k=10$ and $N=5\,000$.} 
\end{figure}

\section{Intrinsic dampening effect}

The reason experiments measure treatment against control in traditional RCTs is to factor out seasonal effects. If the experimenter were to compare the treatment group's before-treatment and during-treatment (or after-treatment) metrics, she would not know whether the measured effect is due to the applied treatment, or something else that changed during the experiment. However, in our models, there is no seasonality, so we can examine $c_{treatment}/c_{base}\prime$. Let's define the intrinsic dampening effect:

\begin{equation}
e_{intrinsic} = c_{treatment} / c_{base}\prime - 1
\end{equation}

This measures the dampening in the effect itself due to the connected, non-independent nature of the network. Naively, we may expect that this ratio should be just the applied treatment effect. However, as we will see, this is not the case. Specifically, for the first experiment considered previously, $e_{intrinsic} = 0.0258$ (with an applied $\Delta \lambda = 0.05$), for the second it is $e_{intrinsic} = 0.0450$ (also with an applied $\Delta \lambda = 0.05$). The full lift is not realized in the treatment group due to content production in the rest of the network being lower, hence, the feedback the treatment nodes receive in their $\nu_i^t = \nu_{damp} \sum_{j \in V_i}{c_j^{t-1}}$ term is lower. If the entire network were to receive the $\Delta \lambda$ boost, in other words if $N=n$, then the measured $e_{intrinsic} = \Delta \lambda / \lambda_{int}$ (which equals $\Delta \lambda$, since $\lambda_{int} = 1$) would hold, as is evident from Equation (2).

\section{Experiment dampening effect}

In an RCT, we measure the treatment versus control group lift, and attribute it to be the effect of the applied treatment, with the caveat that one must be careful to separate out signal from noise using statistical hypothesis testing or bayesian inference. In the terminology of this paper, the treatment effect is defined as:

\begin{equation}
e_{treatment} = c_{treatment} / c_{control} - 1
\end{equation}

In the previous two experiments, we measure $e_{treatment} = 0.0253$ and $e_{treatment} = 0.0449$, respectively. Both are lower than the actual treatment effect $\Delta \lambda = 0.05$ we applied synthetically. This is the \textbf{experiment dampening effect}: because the treatment effect spills over into control and \textit{boosts control as well}, and because treatment nodes do not receive the full network boost (see the previous section), we measure a lower treatment effect. We can define the experiment dampening effect as:

\begin{equation}
e_{dampening} = e_{treatment} / \Delta \lambda
\end{equation}

In the cases above, we get $e_{dampening} = 0.5075$ and $e_{dampening} = 0.8990$, respectively.

Note that the experiment dampening effect and the intrinsic dampening effect are related, but not the same thing:

\begin{itemize}
\setlength\itemsep{0em}
    \item \textbf{experiment dampening effect}: we underestimate the lift in a treatment versus control measurement due to the treatment effect leaking to the control group in the network at treatment time, and due to the intrinsic dampening effect
    \item \textbf{intrinsic dampening effect}: the full lift of the treatment group is not realized in the network due to content production being lower in the rest of the network (the control group) at treatment time; note that the control group does receive some of the treatment lift, so its content production also increases, but not to the level of the treatment group
\end{itemize}

The two are connected by $e_{treatment} \approx e_{intrinsic} - e_{spillover}$. For small treatment groups the intrinsic dampening effect dominates: in the first experiment, $e_{spillover} = 0.0005$ accounts for only a small part of the $0.025$ gap between $e_{treatment}$ and $\Delta \lambda$.

\section{Clustering effect}

In a randomized test, treatment nodes are scattered across the graph, so most of their neighbours are in control; stratification on node characteristics does not change this. The clustering effect is a third effect closely related to the previous two: if the treatment nodes are more densely clustered than the overall graph (or the control group), then the treatment nodes will achieve higher content production, both compared to the control group's $c_{control}$ and the base rate $c_{base}\prime$.

To illustrate the effect, let's look at a Monte Carlo run where the treatment nodes are not randomly selected, but selected to be significantly more clustered. To accomplish tighter clustering in a Watts-Strogatz graph is straightforward: instead of randomly selecting nodes, pick a contiguous set of nodes from the original ring (before randomization). Figure 12 shows such a treatment group on the graph of Figure 1. At the relatively low $p$ edge randomization values that we're using, this will result in a highly clustered subset.

\begin{figure}[h]
	\centering 
	\includegraphics[width=0.34\textwidth]{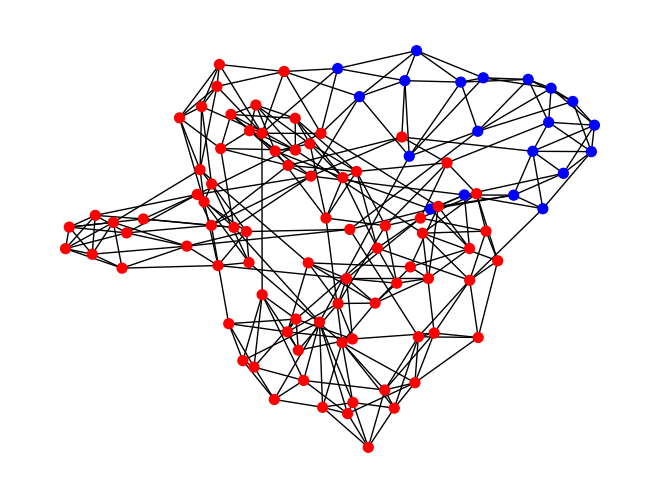}	
	\caption{The same Watts-Strogatz graph from Figure 1, but the treatment group is the first 20 nodes from the original ring.} 
\end{figure}

We run an experiment with identical parameters as in Figure 10, except with a much more clustered treatment group: a Watts-Strogatz graph with $(n=500\,000, k=50, p=0.1)$, $\lambda_{int} = 1$ and $\nu_{damp} = 0.01$, with a 2\% treatment group of $N=10\,000$ nodes receiving a $\Delta \lambda = 0.05$ boost to their content production. Figure 13 shows the resulting content production.

\begin{figure}[h]
	\centering 
	\includegraphics[width=0.34\textwidth]{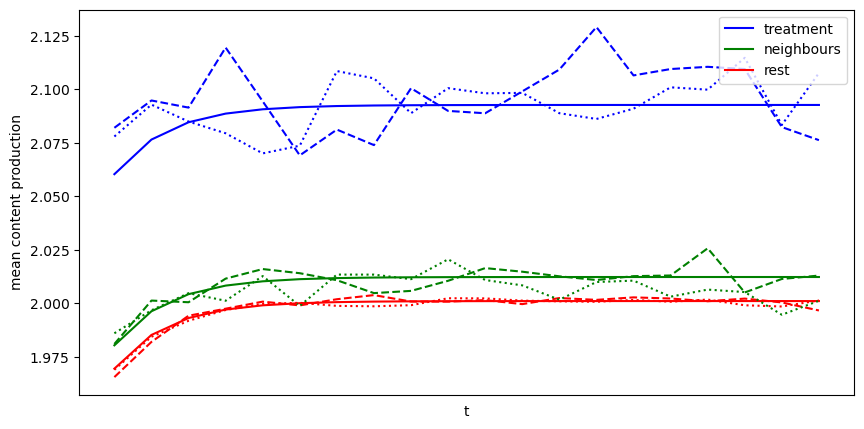}	
	\caption{Same as Figure 10, but the treatment group is tightly clustered, resulting in significantly higher content production.} 
\end{figure}

With a random treatment group, the $10\,000$ treatment nodes had $313\,507$ non-treatment neighbours and about 2\% of treatment nodes' neighbours were also in treatment; now the $10\,000$ treatment nodes only have $46\,545$ non-treatment neighbours and about 90\% of treatment nodes' neighbours are also in treatment. This highly clustered model yields:

\begin{center}
\begin{tabular}{ |c|c|c| } 
 \hline
 \textbf{group} & \textbf{cardinality} & $c / c_{base}\prime$ \\ 
 \hline
 treatment   &  $10\,000$   &  1.0457 \\ 
 control     &  $490\,000$  &  1.0001 \\ 
 neighbours  &  $46\,545$   &  1.0052 \\ 
 rest        &  $443\,455$  &  0.9995 \\ 
 \hline
\end{tabular}
\end{center}

The highly clustered case achieves $e_{dampening} = 0.9136$, compared to $e_{dampening} = 0.5075$ with a random treatment group: the treatment lift is 91\% of the actual $\Delta \lambda$ lift, instead of 51\%. The explanation is that 90\% of the treatment group nodes' neighbours are also in treatment, so they get "their own boost back", and there is relatively little dampening. It is also worth noting that $e_{spillover} = 0.0001$, 5x lower than with a truly random treatment group. The explanation is that there are fewer edges running between treatment and control, so fewer ways for the spillover to happen.

\section{Experiment size effect}

The last effect we consider is the experiment size effect. The experiment size effect is closely related to the spillover effect, and is simply the observation that with a larger treatment group, there is stronger spillover, which yields stronger feedback, which results in higher content production in the treatment group. Figure 14 shows the variation of $e_{intrinsic}$ with $N/n$ for a Watts-Strogatz graph with $n=10\,000, p=0.1$ and $\Delta \lambda = 0.05$.

\begin{figure}[h]
	\centering 
	\includegraphics[width=0.34\textwidth]{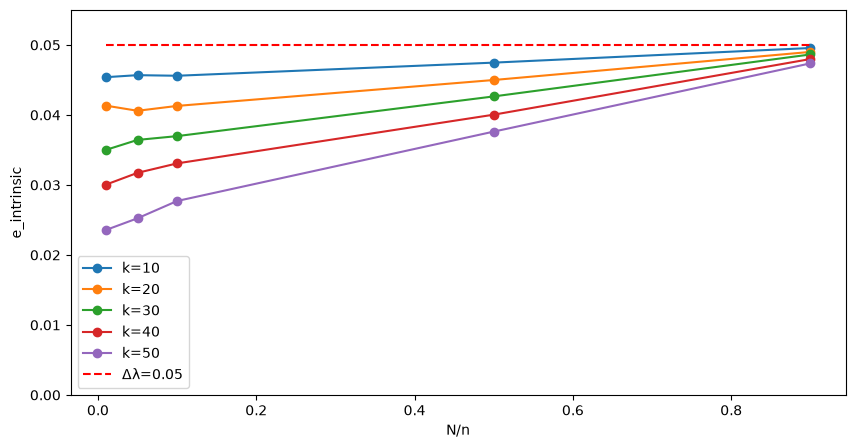}	
	\caption{As $N/n$ approaches 1, the intrinsic effect (the ratio of treatment content production to base content production) approaches the experiment effect $\Delta \lambda$. Experiment ran on a Watts-Strogatz graph with $n=10\,000$ and $p=0.1$, at different $k$ values.} 
\end{figure}

However, as pointed out earlier, in an RCT, the experimenter compares treatment content production to control's (and not base). Figure 15 shows $e_{dampening}$ (the ratio of the treatment effect $e_{treatment}$ to the actual $\Delta \lambda$ effect) for the same experiment. Remarkably, the dampening is essentially constant with $N/n$, which means that even with increasing $N$, the experimenter would measure the same treatment effect, because due to spillover, control's content production also goes up. As $N/n$ grows, the intrinsic dampening effect shrinks and the spillover into control grows, and the two approximately offset each other in $e_{treatment}$.

\begin{figure}[h]
	\centering 
	\includegraphics[width=0.34\textwidth]{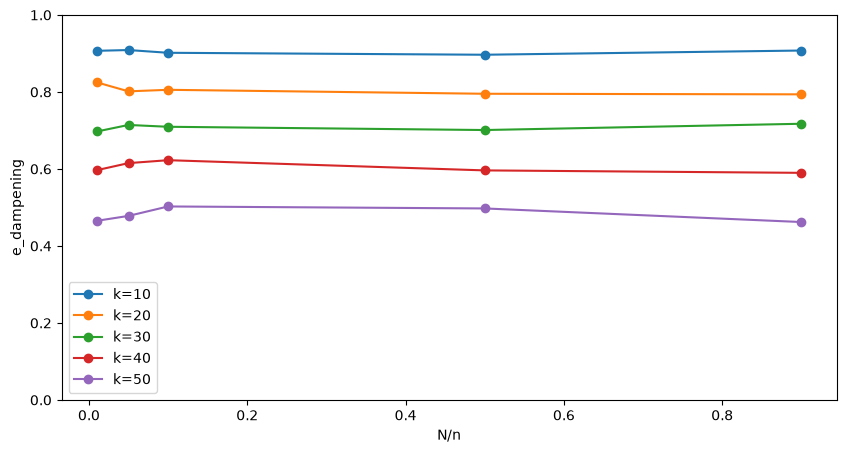}	
	\caption{The dampening effect (the ratio of the treatment effect to $\Delta \lambda$) is approximately constant with $N/n$, and is mostly a function of $k$.} 
\end{figure}

Both plots split as a function of $k$, the mean degree: the lower $k$, the closer we are to the Independence Assumption, the more the measured treatment effect approaches $\Delta \lambda$. Separate experiments (not discussed here) show no dependence on $p$.

\section{Conclusion}

In traditional RCTs where the Independence Assumption holds, the experimenters' main focus is on study design and controlling for statistical fluctuations between the various experiment groups using techniques such as A/A tests, sample ratio mismatch (SRM) tests, p-values, Bayesian testing, and so on. This paper showed that when running experiments on social networks or other graphs, where the IA does not hold, various other effects, all a result of the network effect, must be considered and controlled for. As the last section showed, at any treatment group size, the experimenter will never measure the true treatment effect when comparing treatment to control: in small treatment groups because treatment nodes do not receive the full network boost, in large ones because the treatment effect spills over to the control group. This suggests two remedies. First, the experimenter can estimate the underlying graph structure and the strength of the network effect, and use simulations similar to the ones in this paper to calculate the expected dampening, which converts the measured lift into an estimate of the effect of a full rollout. Second, as the clustering effect shows, selecting treatment nodes in clusters reduces the dampening at the source.

\section{Code}

The paper's accompanying code, including all figures in the paper, is available as a Python notebook on Github. The random seed has been set to a fixed value in the code, so all numerical results are reproducible by re-running the notebook. Use the below link to access the notebook:
\texttt{{\scriptsize https://github.com/mtrencseni/mc-network-effects-rct }}

\bibliographystyle{elsarticle-harv}

\end{document}